\documentstyle[amssymb,amsmath,aps,epsfig,twocolumn]{revtex}

\newcommand{\clebsch}[6]{{\left(
      \begin{array}{cc|c}
        {#1} & {#3} & {#5} \\
        {#2} & {#4} & {#6}
      \end{array}\right)}}
\newcommand{\sixj}[6]{{\left\{
      \begin{array}{ccc}
        {#1} & {#2} & {#3} \\
        {#4} & {#5} & {#6}
      \end{array}\right\}}}

\title{Considerations on the quantum double-exchange Hamiltonian}
\author{A. Wei{\ss}e~$^a$, J. Loos~$^b$, and H.~Fehske~$^a$}
\address{
  $^a$~Physikalisches Institut, Universit\"a{}t Bayreuth, 
  95440 Bayreuth, Germany\\ 
  $^b$~Institute of Physics, Czech Academy of Sciences, 16200
  Prague, Czech Republic\\
  {\rm (\today)}\\[0.5cm]}
\address{~\parbox{14cm}{\rm
    Schwinger bosons allow for an advantageous representation of
    quantum double-exchange. We review this subject, comment on 
    previous results, and address the transition to the semiclassical limit.
    We derive an effective fermionic Hamiltonian for the spin-dependent 
    hopping of holes interacting with a background of local spins, which 
    is used in a related publication within a two-phase description 
    of colossal magnetoresistant manganites. 
    }}

\begin{document}
\maketitle

\section{Introduction}
Introduced by Zener~\cite{Ze51b} in the early 1950s the notion 
of double exchange together with mixed-valency manganites
$R_{1-x}A_x$MnO$_3$ (where $R$ = La, Pr, Nd and $A$ = Sr, Ca, Ba,
Pb) attracted renewed attention when a colossal magnetoresistive effect 
was discovered in these compounds some years ago~\cite{JTMFRC94}. 
The magnetic and electronic properties of manganese oxides,
to some extend, are believed to arise from the large Coulomb and Hund's
rule interaction of the manganese $d$ shell electrons. Due to the
almost octahedral coordination within the perovskite structure the
$d$ levels split into two subbands labeled according to their
octahedral symmetry, $e_g$ and $t_{2g}$. In the case of zero doping
($x=0$) there are four electrons per Mn site which fill up the three
$t_{2g}$ levels and one $e_g$ level, and by Hund's rule, form a $S=2$
spin state. Doping will remove the electron from the $e_g$ level, and
by hopping via bridging oxygen sites these holes acquire mobility.
However, this hopping acts in a background of local spins $S=3/2$ formed
by the $t_{2g}$ electrons and its amplitude depends on the overlap of 
the spin states at neighbouring sites (or, in a classical language, on 
their relative angle), it is largest if the total bond spin is maximal 
and vice versa~\cite{AH55}.

Another ingredient, that is assumed to significantly influence the 
physical properties of manganites, is electron-lattice interaction. 
Namely the two $e_g$ orbitals, which are degenerate in a perfect
cubic environment, will couple to lattice vibrations of the same
symmetry, giving rise to a Jahn-Teller effect and polaronic behaviour
in some regions of the phase diagram.
It is this close interplay of three different subsystems (electrons in
degenerate orbitals, background of localized spins, and lattice vibrations) 
that makes the physics of manganites both, rich and
complicated. 

In the present work we concentrate on the double exchange (DE)
part of the interactions and consider different possibilities for an
approximate treatment of the exact DE Hamiltonian on a lattice in terms of
effective electronic one- or two-band models. These can be used in a more
elaborate modelling of the real materials (see our forthcoming 
work~\cite{WLF01b}). 
It turns out that quantum double exchange on a lattice is most suitably 
derived and described with the help of Schwinger bosons. We therefore 
include a detailed and pedagogic derivation of the quantum DE Hamiltonian 
using Schwinger bosons.
Although this approach has been used before~\cite{Ki99,GWSX00}, 
we feel a comprehensive presentation of the subject is still missing. 
In two appendices we reexamine the derivation for two sites, and consider 
the semiclassical limit ($S\rightarrow\infty$). In addition, by means 
of numerical experiments, we illustrate how this limit evolves from 
the quantum case.

\section{Schwinger boson representation of double exchange}

To derive the quantum DE Hamiltonian on a lattice, as a starting point
we take the Kondo lattice model including on-site Coulomb repulsion,
\begin{eqnarray}\label{hstart}
  H  & = & -t \sum_{\langle ij\rangle \sigma} 
  \left[c_{i\sigma}^{\dagger} c_{j\sigma}^{} + \textrm{H.c.}\right] 
  - J_H \sum_{i \sigma \sigma'} 
  ({\bf S}_i \bbox{\sigma}_{\sigma\sigma'}) 
  c_{i\sigma}^{\dagger} c_{i\sigma'}^{}\nonumber{}\\
  & & +\ U\sum_i n_{i\downarrow} n_{i\uparrow}\,,
\end{eqnarray}
where summation is over nearest neighbour bonds $\langle ij\rangle$
or sites $i$, respectively.
For clarity and since it can be included easily in the final result, 
here we have neglected the orbital degeneracy of the $e_g$ electrons. 
That is, $c_{i\sigma}^{(\dagger)}$ denote electrons
in a single band, which interact with some localized spin ${\bf S}_i$
via the Hund's coupling $J_H$. In the real materials this localized spin
corresponds to the remaining $t_{2g}$ electrons, that tend to form
a high spin state with an electron in the $e_g$ shell.

In the manganites the situation is such, that $U\gg J_H>t$ (cf. 
Refs.~\cite{AH55,BMSNF92}
). Hence, we first take the limit $U\rightarrow\infty$, resulting in
\begin{equation}
\label{huu}
  H = -t \sum_{\langle ij\rangle \sigma} 
  \left[\tilde c_{i\sigma}^{\dagger} \tilde c_{j\sigma}^{} 
    + \textrm{H.c.}\right] 
  - J_H \sum_{i \sigma \sigma'} 
  ({\bf S}_i \bbox{\sigma}_{\sigma\sigma'}) 
  \tilde c_{i\sigma}^{\dagger} \tilde c_{i\sigma'}^{}
\end{equation}
with restricted fermions $\tilde{c}_{i\sigma}=c_{i\sigma}(1-n_{i\,-\sigma})$.
Next, following Kubo and Ohata~\cite{KO72a}, the exchange term in 
Eq.~(\ref{huu}) is solved while the hopping term is considered as 
a small perturbation.
For positive $J_H$ the ground-state of the exchange term is a free 
spin $S$, if there is no electron at site $i$, or a coupled spin 
$\bar S = S+1/2$ otherwise (note, that we use $S$ for the 
length of the localized spin ${\bf S}_i$ formed by $t_{2g}$ electrons).  
To describe the effective hopping we therefore need a projection 
operator, which restores these conditions,
\begin{equation}
\label{poe1}
  (P_i^+)_{\sigma\sigma'} = 
  \frac{({\bf S}_i \bbox{\sigma}_{\sigma\sigma'}) 
    + (S+1)\delta_{\sigma\sigma'}}{2S+1}\,.
\end{equation}
Then the DE-Hamiltonian (Eq.~(2.3) in Ref.~\cite{KO72a}) 
is given in terms of spin and restricted fermion operators by
\begin{equation}
  \label{hde}
  H^{\rm DE}_{\rm el} = -t \sum_{\langle ij\rangle\sigma\sigma'} 
  \left[\tilde c_{i\sigma}^{\dagger} (P_i^+ P_j^+)_{\sigma\sigma'}^{}
  \tilde c_{j\sigma'}^{} + \textrm{H.c.}\right]\,. 
\end{equation}
However, this expression turns out to be unwieldy for analytic as 
well as numeric calculations. Although in principle the electronic
spin is absorbed into the total spin at each site, the spin index
$\sigma$ is still present in Eq.~(\ref{hde}). Here the advantages
of Schwinger bosons come into play, namely the possibility to 
describe spins of arbitrary amplitude with the same set of 
boson operators $a_i$ and $b_i$,
\begin{eqnarray}
  \label{sb1}
  S_i^+ & = & a_i^{\dagger} b_i^{}\,,\  S_i^- =  b_i^{\dagger} a_i^{}\,,\\
  \label{sb2}
  S_i^z & = & (a_i^{\dagger} a_i^{} - b_i^{\dagger} b_i^{})/2\,,\\
  \label{sb3}
  |S_i| & = & (a_i^{\dagger} a_i^{} + b_i^{\dagger} b_i^{})/2\,.
\end{eqnarray}
Using these operators, we can rewrite the projection operators $P_i^+$,
\begin{eqnarray}
  \label{poe2}
  (P_i^+)_{\sigma\sigma'}^{} & = & \frac{1}{2S+1}\left[\begin{array}{cc}
      (S+1) + S_i^z & S_i^-\\
      S_i^+ & (S+1) - S_i^z
    \end{array}\right]\nonumber{}\\
  & = & \frac{1}{2S+1}\left[\begin{array}{cc}
      a_i^{} a_i^{\dagger} & a_i^{} b_i^{\dagger}\\
      b_i^{} a_i^{\dagger} & b_i^{} b_i^{\dagger}
    \end{array}\right]\,,
\end{eqnarray}
where we can keep $S$ in the denominator, because it is conserved.
The last matrix can be decomposed easily,
\begin{equation}
  \label{poe3}
  (P_i^+)_{\sigma\sigma'}^{}  =  \frac{1}{2S+1}
  \left[\begin{array}{c} a_i^{} \\ b_i^{} \end{array}\right]\cdot
  \left[\begin{array}{cc} a_i^{\dagger} & b_i^{\dagger} \end{array}\right]\,,
\end{equation}
which leads to
\begin{equation}\label{hboson}
  H^{\rm DE}_{\rm el}  = \frac{-t}{2S+1} \sum_{\langle ij\rangle}
  \left[(R_i^+)^{\dagger} (a_i^{\dagger} a_j^{} + b_i^{\dagger} b_j^{})R_j^{+}+
  \textrm{H.c.}\right] 
\end{equation}
with the projectors
\begin{equation}
  \label{roe}
  R_i^+ = \frac{\tilde c_{i\uparrow} a_i^{\dagger}
    +\tilde c_{i\downarrow} b_i^{\dagger}}{\sqrt{2S+1}}\,.
\end{equation}
If we restrict the Hilbert space of the problem by fixing the spin
length at each site to the value $S+n_i/2$ (where $n_i = 
\tilde n_{i\uparrow} + \tilde n_{i\downarrow}$), we can simply
replace the projectors $R_i^+$ by spinless fermion operators $c_i$.
This can be seen, by taking a closer look on the operators $R_i^+$.
Let us assume, site $i$ is occupied by an electron, and therefore the 
total spin at the site is $S+1/2$.
The corresponding state $|n_e;S,m\rangle$ can be written as
\begin{eqnarray}\label{elspin}
  |1;S+\tfrac{1}{2},m\rangle & = &
  \sqrt{\frac{S+\frac{1}{2}+m}{2S+1}}\ 
  |\!\uparrow\rangle |S,m-\tfrac{1}{2}\rangle \nonumber{}\\
  & + & \sqrt{\frac{S+\frac{1}{2}-m}{2S+1}}\ 
  |\!\downarrow\rangle |S,m+\tfrac{1}{2}\rangle\,.
\end{eqnarray}
Using the representation of an $S^z$ eigenstate in terms of Schwinger bosons,
\begin{equation}
  \label{sws}
  |S,m\rangle = 
  \frac{(a^{\dagger})^{S+m}(b^{\dagger})^{S-m}}{\sqrt{(S+m)!(S-m)!}}
  \ |0\rangle\,,
\end{equation}
we find, that applying the operator $R^+$ the state
$|1;S+1/2,m\rangle$ in Eq.~(\ref{elspin}) is transformed
into the corresponding Schwinger boson representation 
of the coupled spin $S+1/2$, while the electron is
annihilated
\begin{equation}
  \label{ros}
  R^+ |1;S+\tfrac{1}{2},m\rangle = |0;S+\tfrac{1}{2},m\rangle\,.
\end{equation}
Backwards, the operator $(R^+)^{\dagger}$ creates the decomposition 
of the coupled spin into electronic and localized spin, i.e., it 
produces appropriate Clebsch-Gordan coefficients.

Now it is straightforward to omit the electronic spin index, using
only spinless fermions, $c_i^{(\dagger)}$, and Schwinger bosons to 
describe double-exchange. The corresponding Hamiltonian is given by
\begin{equation}\label{hfinal}
  H^{\rm DE}_{\rm el} = \frac{-t}{2S+1}\sum_{\langle ij\rangle} \left[
    (a_i^{\dagger} a_j^{} + b_i^{\dagger} b_j^{})
    c_{i}^{\dagger} c_{j}^{} + \textrm{H.c.}\right]\,,
\end{equation}
where, for every site $i$, the Hilbert space is constrained to
\begin{equation}
  \label{co1}
  a_i^{\dagger} a_i^{} + b_i^{\dagger} b_i^{} = 2 S + c_i^{\dagger} c_i^{}\,.
\end{equation}

In the case of low doping usually it is more appropriate and natural 
to consider holes instead of electrons. Here ``hole'' denotes a fermion
and a spin $S$ moving together in a background of spins $\bar S = S+1/2$. 
For the transformation of the electronic Hamiltonian, Eq.~(\ref{hboson}), 
the operator $R^+$ needs to be replaced by a suitable counterpart 
involving restricted hole operators $\tilde h_{i\sigma}$.
In analogy to Eqs.~(\ref{elspin}) to~(\ref{ros}), the state
\begin{eqnarray}
  \label{hospin}
  |n_h=1;S,m\rangle & = & 
  \sqrt{\frac{\bar S-m+\frac{1}{2}}{2\bar S+1}}\ 
  |\!\uparrow\rangle |\bar S,m-\tfrac{1}{2}\rangle \nonumber{}\\
  & - & \sqrt{\frac{\bar S+m+\frac{1}{2}}{2\bar S+1}}\ 
  |\!\downarrow\rangle |\bar S,m+\tfrac{1}{2}\rangle
\end{eqnarray}
is transformed into its corresponding Schwinger boson representation
by the operator
\begin{equation}
  \label{roh} 
  R_i^- = \frac{\tilde h_{i\uparrow} b_{i} - 
    \tilde{h}_{i\downarrow} a_{i}}{\sqrt{2 {\bar S} + 1}}\,.
\end{equation}
Hence, in hole representation, the DE Hamiltonian reads
\begin{equation}
  \label{hdeho}
  H_{\rm hole}^{\rm DE} = \frac{t}{2\bar{S}} \sum_{\langle ij\rangle}
  \left[(R_i^-)^{\dagger}  (a_i^{} a_j^{\dagger} + b_i^{} b_j^{\dagger}) 
    R_j^{-} + \textrm{H.c.}\right]\,.
\end{equation}
The Hamiltonian for spinless fermions, Eq.~(\ref{hfinal}), 
changes only little, becoming
\begin{equation}
  \label{hfinal2}
  H_{\rm hole}^{\rm DE} = \frac{t}{2\bar S}\sum_{\langle ij\rangle} 
  \left[(a_i^{} a_j^{\dagger} + b_i^{} b_j^{\dagger})
    h_{i}^{\dagger} h_{j}^{} + \textrm{H.c.}\right]
\end{equation}
with the constraint
\begin{equation}
  \label{co2}
  a_i^{\dagger} a_i^{} + b_i^{\dagger} b_i^{} = 
  2 \bar S - h_i^{\dagger} h_i^{}\,.
\end{equation}

\section{Effective transport Hamiltonian}
To obtain an effective Hamiltonian for the spin-dependent hole-hopping,
the spin part of the DE interaction is considered within mean field
approximation. However, there are two representations of the DE
Hamiltonian to start from: Eqs.~(\ref{hdeho}) and~(\ref{hfinal2}).
The resulting effective Hamiltonians describe carriers with or without 
spin, respectively.
Given Eq.~(\ref{hfinal2}), a mean hopping of the spinless carriers, 
$\tilde{t}^{\rm (b)}$, is obtained by considering each 
bond~$\langle ij \rangle$ separately, and taking the expectation value 
of the DE term for all values of the total bond spin $S_T$ and $S_T^z$
in an effective ordering field $\lambda = \beta g\mu_B H^z_{\rm eff}$.
This reproduces the result of Kubo and Ohata~\cite{KO72a}.
Having spin $\bar S$ at site $i$ and spin $S$ at site $j$, the coupled 
state $S_T$ with maximal $S_{T}^{z}$ is given by
\begin{eqnarray}
  \label{smst}
  \lefteqn{|S_T, S_T\rangle_{(\bar S S)} =}\nonumber{}\\
  & = & C\ (b_{i}^{\dagger} a_{j}^{\dagger} - 
  b_{j}^{\dagger} a_{i}^{\dagger})^{2S+\frac{1}{2}-S_T} 
  (a_{i}^{\dagger})^{S_T+\frac{1}{2}} 
  (a_{j}^{\dagger})^{S_T-\frac{1}{2}} |0\rangle\,,
\end{eqnarray}
where $C$ is a normalization factor,
\begin{equation}
  \label{c}
  C = \sqrt{\frac{(2 S_T + 1)!}{(S_T+\frac{1}{2})!(S_T-\frac{1}{2})!
      (2 S+\frac{1}{2}-S_T)!(2 S+\frac{3}{2}+S_T)!}}\,.
\end{equation}
Applying $(a_j^{\dagger} a_i^{} + b_j^{\dagger} b_i^{})$ we arrive at
\begin{eqnarray}
  \label{osmst}
  \lefteqn{(a_{j}^{\dagger} a_{i}^{} + b_{j}^{\dagger} b_{i}^{})
    |S_T, S_T\rangle_{(\bar S S)} = }\nonumber{}\\
  & = & C (S_T + \tfrac{1}{2}) 
  (b_{i}^{\dagger} a_{j}^{\dagger} - 
  b_{j}^{\dagger} a_{i}^{\dagger})^{2S+\frac{1}{2}-S_T} 
  (a_{i}^{\dagger})^{S_T-\frac{1}{2}} 
  (a_{j}^{\dagger})^{S_T+\frac{1}{2}} |0\rangle\nonumber{}\\
  & = &  (S_T + \tfrac{1}{2}) |S_T, S_T\rangle_{(S \bar S)}\,.
\end{eqnarray}
Hence, we rederived the effective matrix element for a 
single bond (cf. Appendix~\ref{app2site} or Ref.~\cite{AH55}),
\begin{equation}
  \label{tb}
  t^{\rm (b)} = \frac{S_T + 1/2}{2\bar S}\ t\,,
\end{equation}
which is averaged over all values and directions of $S_T$,
\begin{equation}
  \label{ttilde}
  \tilde{t}^{\rm (b)} = \gamma_{\bar S}[\bar S\lambda]\ t\,,
\end{equation}
where~\cite{KO72a}
\begin{eqnarray}
  \label{gsb}
  \lefteqn{\gamma_{\bar S}[\bar S \lambda] =  
    \frac{\sum_{S_T = 1/2}^{2\bar S - 1/2} 
      \sum_{M=-S_T}^{S_T} \frac{S_T+1/2}{2\bar S} \mbox{ e}^{M \lambda}}
    {\sum_{S_T = 1/2}^{2\bar S - 1/2}
      \sum_{M=-S_T}^{S_T} \mbox{ e}^{M \lambda}}}\\
  & = & \tfrac{1}{2} + \tfrac{\bar S}{2\bar S + 1}
  \coth\!\left(\tfrac{2\bar S + 1}{2}\lambda\right)
  \left[\coth(\bar{S}\lambda) - \tfrac{1}{2\bar S} 
    \coth\!\left(\tfrac{\lambda}{2}\right)\right]\,.\nonumber{}
\end{eqnarray}
The effective Hamiltonian describing spinless fermionic holes in an 
averaged background of ordered spins reads 
\begin{equation}
  \label{hhoeffI}
  H_{\rm hole}^{\rm eff, I} = \tilde{t}^{\rm (b)} \sum_{\langle ij\rangle}
  \left[ h_{i}^{\dagger} h_{j}^{} + \textrm{H.c.}\right]\,.
\end{equation}
In Appendix~\ref{appclass} we compare the classical limit of both, 
this Hamiltonian and the exact expression. At least the bandwidth turns 
out to be represented very well.

Another possible way to obtain an effective Hamiltonian is based on the
picture of itinerant carriers of spin $\frac{1}{2}$ moving in the background
of localized spins, the correlations of which change on a large time scale
compared with the hopping frequency. Then an effective hopping Hamiltonian
is obtained averaging $H_{\rm hole}^{\rm DE}$, Eq.~(\ref{hdeho}), 
over free spins $\bar S$ in a homogeneous field $\lambda$. According to 
Eqs.~(\ref{sb1})-(\ref{sb2}) and the fact that $\langle S^{\pm}\rangle = 0$ 
and $\langle S^{z}\rangle = \bar{S} B_{\bar S}[\bar{S}\lambda]$ (where 
$B_{\bar S}[z]$ denotes the Brillouin function), only two terms contribute.
The resulting Hamiltonian involves two effective hopping matrix
elements, one for each spin channel,
\begin{equation}
  \label{hhoeffII}
  H_{\rm hole}^{\rm eff, II} = 
  \sum_{\langle ij\rangle}\left[
    \tilde{t}_{\uparrow}\tilde{h}_{i\uparrow}^{\dagger}
    \tilde{h}_{j\uparrow}^{} + 
    \tilde{t}_{\downarrow}\tilde{h}_{i\downarrow}^{\dagger} 
    \tilde{h}_{j\downarrow}^{} + 
    \textrm{H.c.}\right]
\end{equation}
with
\begin{eqnarray}
  \label{ttildeu}
  \tilde t_{\uparrow} & = & \frac{t}{(2\bar S)(2\bar S +1)}
  \left[\bar S (1 - B_{\bar S}[\bar{S}\lambda])\right]^2\,,\\
  \tilde t_{\downarrow} & = & \frac{t}{(2\bar S)(2\bar S +1)}
  \left[\bar S (1 + B_{\bar S}[\bar{S}\lambda])\right]^2\,.
  \label{ttilded}
\end{eqnarray}
In a fully polarized background ($\lambda\to\infty$) only holes with 
anti-parallel spin can hop 
($\tilde t_{\downarrow}\to\frac{2\bar S}{2\bar S+1}\ t$), 
while holes with parallel spin are blocked ($\tilde t_{\uparrow}\to 0$).
In general, the situation is complicated by the fact that Eq.~(\ref{hhoeffII})
involves restricted fermion operators (Hubbard operators) forbidding 
an exact solution of the model. However, to a good approximation, in the 
polarized phase the up-band can be neglected, while the down-band is taken 
into account using unrestricted operators $h_{i\downarrow}^{(\dagger)}$.
On the other hand, in a disordered phase ($\lambda\to 0$) both bands are
equivalent 
($\tilde t_{\uparrow}=\tilde t_{\downarrow}\to\frac{\bar S/2}{2\bar S+1}\ t$),
making an approximate treatment less evident.

\section{Conclusions}
In the present work we review the subject of double exchange and
derive an effective Hamiltonian for the spin-dependent hopping 
of holes in an averaged background of local spins. This is used 
in a subsequent publication within a two-phase scenario for the description 
of colossal magnetoresistant manganites, alternatively to the effective 
Hamiltonian for spinless carriers derived in Ref.~\cite{KO72a}.
Besides, we illustrate that on the level of quantum spins all features 
of double exchange - namely its derivation, limiting cases, as well
as approximations - are probably most clearly represented in terms
of Schwinger bosons. 

{\it Acknowledgements:}
This work was supported by the Deutsche Forschungsgemeinschaft
and the Czech Academy of Sciences under Grant No. 436 TSE 113/33.

\appendix
\section{Reexamination of the 2-site problem}\label{app2site}
As noted above, the DE matrix element, Eq.~(\ref{tb}), was first
derived by Anderson and Hasegawa~\cite{AH55} by considering a
system of two spins $S$ at neighbouring sites and a mobile electron 
whose spin $\sigma$ is coupled to the local spin at the same site.
Recently the problem was reexamined by M{\"u}ller-Hartmann and
Dagotto~\cite{MD96}, who found a ``nontrivial'' phase factor, which
the authors interpreted in terms of a Berry phase within the 
limit $S\to\infty$. We argue that the phase factor of their quantum
DE Hamiltonian merely compensates for the phase introduced by the
permutation of neighbouring spin states, whereas independently 
the Berry phase evolves in the classical ($S\to\infty$) limit of the
correct DE Hamiltonian, Eq.~(\ref{hfinal}). Moreover, on a lattice
double exchange can not be described in terms of spin and permutation
operators, which do not take into account fermionic commutation
relations.

To be specific we briefly summarize the two site problem.
Assuming the electron initially to be at site~1 we start with 
coupling the local spin ${\bf S}_1$ and the 
electron spin $\bbox{\sigma}$, to give an on-site spin 
$\bar{\bf S}_1 = {\bf S}_1 + \bbox{\sigma}$. We note, that the
construction of the corresponding wave function is not unique,
as is known from textbooks on spin algebra~\cite{Li84}. In more detail
\begin{eqnarray}
  |\bar S_1\bar m_1\rangle_{(\sigma S_1)} & = & \sum_{\mu m_1} 
  \clebsch{\sigma}{\mu}{S_1}{m_1}{\bar S_1}{\bar m_1}
  |\sigma\mu\rangle|S_1m_1\rangle\,,\\
  |\bar S_1\bar m_1\rangle_{(S_1\sigma)} & = & 
  (-1)^{\sigma+S_1-\bar S_1} |\bar S_1\bar m_1\rangle_{(\sigma S_1)} \,,
\end{eqnarray}
where we give the order of the constituting spin states as a
subscript and denote Clebsch-Gordan coefficients by 
$(\cdot\cdot\!\mid\!\cdot)$. 
Adding this spin to the unchanged core spin at site~2, we 
arrive at an initial state of total spin $S_T$
\begin{eqnarray}
  |\varphi_{\rm in}^{}\rangle & = & |S_T m_T\rangle_{((\sigma S_1)S_2)}\\
  & = & \sum_{\bar m_1 m_2} 
  \clebsch{\bar S_1}{\bar m_1}{S_2}{m_2}{S_T}{m_T}
  |\bar S_1 \bar m_1\rangle_{(\sigma S_1)} |S_2 m_2\rangle\,.
\end{eqnarray}
Now, if the electron moves to site~2, we can couple $\bbox{\sigma}$ and
${\bf S}_2$ to give $\bar{\bf S}_2$, which together with
${\bf S}_1$ yields another state of total spin $S_T$.
However, as can be seen above, there are different ways
to connect the three spins. One possible final state is
\begin{eqnarray}
  |\varphi_{\rm fi}^A\rangle & = & |S_T m_T\rangle_{(S_1(\sigma S_2))}\\
  & = & \sum_{m_1 \bar m_2} 
  \clebsch{S_1}{m_1}{\bar S_2}{\bar m_2}{S_T}{m_T}
  |S_1 m_1\rangle |\bar S_2 \bar m_2\rangle_{(\sigma S_2)}\,.
\end{eqnarray}
The corresponding effective hopping matrix element $t^{\rm (b)}_A$
is proportional to the overlap $\langle \varphi_{\rm fi}^A|\varphi_{\rm in}^{}\rangle$, which 
can be evaluated by rewriting the multiple sum over the 
product of four Clebsch-Gordan coefficients with a 6-j-symbol,
\begin{eqnarray}
  \lefteqn{\langle \varphi_{\rm fi}^A|\varphi_{\rm in}^{}\rangle\ =}\\ 
  & &  (-1)^{\bar S_1 + S_2 + S_T} 
  \sqrt{(2\bar S_1 +1)(2 \bar S_2 +1)} 
  \sixj{\sigma}{S_1}{\bar S_1}{S_T}{S_2}{\bar S_2}\nonumber{}\,.
\end{eqnarray}
We obtain the value of the matrix relevant for strong Hund's coupling
by setting $|\bar{\bf S}_i| = S_i + 1/2 \equiv \bar S$ and 
$|{\bf S}_i| = S = \bar S-1/2$:
\begin{equation}
  t^{\rm (b)}_A = \frac{S_T + 1/2}{2\bar S}\ t\,,
\end{equation}
i.e., there is no $S_T$-dependent phase factor. However, 
M\"u{}ller-Hartmann and Dagotto derive an effective 
double-exchange Hamiltonian, where the hopping is expressed
as the permutation of the spin $\bar S$ ``particle'' at site~1 and
the spin $(\bar S-1/2)$ ``hole'' at site~2. Therefore the final state
they have to consider for the matrix element is
\begin{eqnarray}
  |\varphi_{\rm fi}^B\rangle & = & |S_T m_T\rangle_{((\sigma S_2)S_1)}\\
  & = & \sum_{\bar m_2 m_1} 
  \clebsch{\bar S_2}{\bar m_2}{S_1}{m_1}{S_T}{m_T}
  |\bar S_2 \bar m_2\rangle_{(\sigma S_2)}|S_1 m_1\rangle \,.
\end{eqnarray}
Obviously the permutation of the indices in the Clebsch-Gordan
coefficients yields a different phase factor in the overlap
\begin{eqnarray}
  \lefteqn{\langle \varphi_{\rm fi}^B|\varphi_{\rm in}^{}\rangle\ = }\\
  & & (-1)^{S_1 + S_2 + \bar S_1 + \bar S_2 } 
  \sqrt{(2\bar S_1 +1)(2 \bar S_2 +1)} 
  \sixj{\sigma}{S_1}{\bar S_1}{S_T}{S_2}{\bar S_2}\nonumber{}\,,
\end{eqnarray}
and consequently the authors obtain a $S_T$-dependent phase
factor for $t^{\rm (b)}_B$,
\begin{equation}
  t^{\rm (b)}_B = (-1)^{2\bar S-S_T-1/2}\ \frac{S_T + 1/2}{2\bar S}\ t\,.
\end{equation}
To recover the effective Hamiltonian Eq.(6) of Ref.~\cite{MD96},
\begin{equation}\label{heffmd}
  H^{\rm eff} = -t\ P_{12} Q_{\bar S}(y)\,,
\end{equation}
we simply have to express $S_T$ in $t^{\rm (b)}_B$ with the help of
$y = {\bf S}_1\cdot{\bf S}_2 / ({\bar S} (\bar S -1/2))$. 
In a direct way we can 
use interpolation polynomials in the form of Lagrange, i.e.
\begin{eqnarray}
  Q_{\bar S} (y) & = & \sum_{S_T = 1/2}^{2\bar S-1/2} t^{\rm (b)}_B 
  \prod_{\genfrac{}{}{0pt}{}{j = 1/2}{j\ne S_T}}^{2\bar S-1/2} 
  \frac{y - y_j}{y_{S_T} - y_j}\\
  y_j & = & \frac{j(j+1) - \bar S(\bar S+1) - (\bar S-1/2)(\bar S+1/2) 
    }{ 2\bar S(\bar S-1/2)}\,,
\end{eqnarray}
or the recursive formula given in Eq.(7) of Ref.~\cite{MD96}. 
Of course, taking the matrix element $t^{\rm (b)}_A$ for the
construction of $Q_{\bar S}(y)$ is wrong, as this does not account for
the phase factor due to the permutation $P_{12}$. Concerning the
limit of classical spins, we can see no connection between the 
above $S_T$-dependent phase factor and the Berry phase.

The peculiarities concerning the phase factor indicate that the
above derivation is not suitable for the generalization to a lattice.
Moreover, the spin-$\bar S$ ``particles'' still obey fermion commutation
relations, which can not be expressed by conventional permutation
or spin operators. That means, the expression given in Eq.(6) 
of Ref.~\cite{MD96},
\begin{equation}
  H^{\rm eff} = -t \sum_{\langle ij\rangle} P_{ij} Q_{\bar S}(y)\,,
\end{equation}
is {\it not} the correct quantum DE-Hamiltonian on a lattice.

\section{Classical limit of the DE model}\label{appclass}
The limit $S\rightarrow\infty$ of $H_{\rm el}^{\rm DE}$, Eq.~(\ref{hfinal}), 
is easily derived by taking its expectation value with spin coherent 
states~\cite{Au94},
\begin{equation}
  |\Omega(S,\theta,\phi)\rangle 
  = \frac{(u a^{\dagger} + v b^{\dagger})^{2S}}{
    \sqrt{(2S)!}}\ |0\rangle\,,
\end{equation}
where $u = \cos(\theta/2) \mbox{ e}^{i\phi/2}$ and 
$v = \sin(\theta/2) \mbox{ e}^{-i\phi/2}$.
Using the properties of coherent states,
\begin{eqnarray}
  a\ |\Omega(S,\theta,\phi)\rangle & = & 
  \sqrt{2S}\ u\ |\Omega(S-\tfrac{1}{2},\theta,\phi)\rangle\\
  b\ |\Omega(S,\theta,\phi)\rangle & = & 
  \sqrt{2S}\ v\ |\Omega(S-\tfrac{1}{2},\theta,\phi)\rangle\,,
\end{eqnarray}
for a given spin configuration $\{\theta_k,\phi_k\}$ and 
two electronic states $|\psi_1\rangle$ and $|\psi_2\rangle$,
\begin{equation}
  |\psi_j\rangle = \prod_k 
  |n_{j,k}\rangle|\Omega(S+\tfrac{n_{j,k}}{2},\theta_k,\phi_k)\rangle
\end{equation}
(where $|n_{j,k}\rangle = (c_{k}^{\dagger})^{n_{j,k}}|0\rangle$ with 
numbers $n_{j,k}\in \{0,1\}$), we find the average  
\begin{equation}
  \langle\psi_1| H_{\rm el}^{\rm DE} |\psi_2\rangle = 
  \prod_k \langle n_{1,k}| 
  \Big(-\sum_{\langle ij\rangle} 
    \left[t_{ij}^{} c_{i}^{\dagger} c_{j}^{} + \textrm{H.c.}\right]\Big) 
  \prod_k |n_{2,k}\rangle
\end{equation}
with the matrix element
\begin{eqnarray}
  t_{ij}^{} & = & \cos\left(\frac{\theta_i}{2}\right)
  \cos\left(\frac{\theta_j}{2}\right) \mbox{ e}^{-i(\phi_i-\phi_j)/2}
  \nonumber{}\\
  & + & \sin\left(\frac{\theta_i}{2}\right)
  \sin\left(\frac{\theta_j}{2}\right) \mbox{ e}^{i(\phi_i-\phi_j)/2}\,.
\end{eqnarray}
Hence, the classical Hamiltonian should read
\begin{equation}\label{hclass}
  H_{\rm class}^{\rm DE} = 
  -\sum_{\langle ij\rangle} 
  \left[t_{ij}^{} c_{i}^{\dagger} c_{j}^{} + \textrm{H.c.}\right]\,,
\end{equation}
which is equivalent to the results obtained in Refs.~\cite{KA88,MD96}.
In our case, however, the classical limit followed from the quantum
Hamiltonian, Eq.~(\ref{hfinal}), in an obvious and more straightforward way.
Note also, that averaging the operators $(R_i^+)^{\dagger}$ and $R_i^+$ 
over coherent states yields the unitary transformation
onto rotated electrons $d_j^{(\dagger)}$ (compare Refs.~\cite{KA88,MD96}),
\begin{equation}
  d_j = \cos\left(\frac{\theta_j}{2}\right) \mbox{ e}^{-i\phi_j/2}\ 
  \tilde c_{j\uparrow}^{} +
  \sin\left(\frac{\theta_j}{2}\right) \mbox{ e}^{i\phi_j/2}\ 
  \tilde c_{j\downarrow}^{}\,,
\end{equation}
i.e., naturally Eq.~(\ref{hboson}) has the same classical limit.

\begin{figure}[htb]
  \epsfig{file=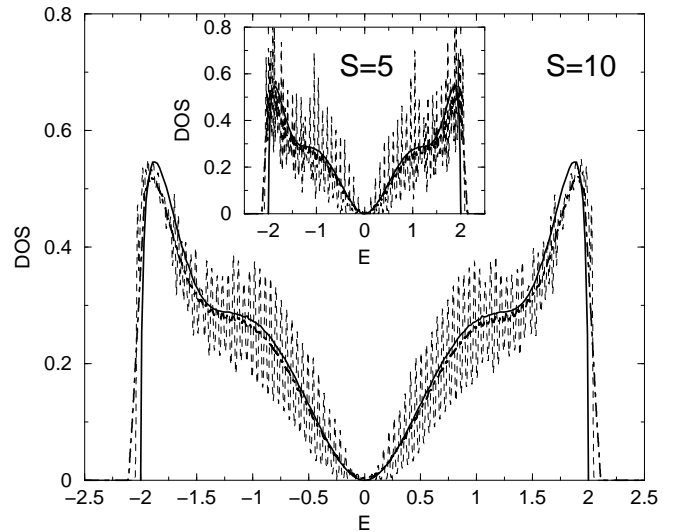, width=\linewidth}
  \caption{Density of nonzero eigenvalues (dashed line) and 
    running average (bold dot-dashed), calculated for 2 electrons 
    on 4 sites with $S=5$ (inset) and $S=10$, compared to the 
    classical result $S\rightarrow\infty$ (bold solid).}\label{figdosn4}
\end{figure}

To check, whether the description of double-exchange in terms of classical
spins is appropriate, we compared the (canonical) density of states (DOS)
for a fixed number of carriers on a small cluster, which interact with
quantum (Eq.~(\ref{hfinal})) or classical (Eq.~(\ref{hclass})) spins, 
respectively. 
Using Chebychev expansion and maximum entropy 
methods~\cite{SRVK96}, 
for the quantum case the spectra can be obtained numerically for rather 
large $S$. The classical DOS is found by averaging the eigenvalues of 
$H_{\rm class}^{\rm DE}$, Eq.~(\ref{hclass}), over a large number of 
spin configurations. 
\begin{figure}[htb]
  \epsfig{file=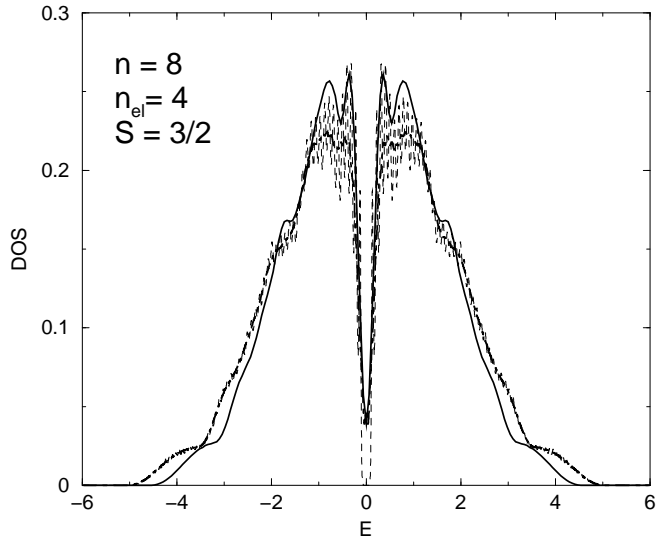, width=\linewidth}
  \caption{Density of nonzero eigenvalues for 4 electrons on 8 sites
    with $S=3/2$; same notations as in Fig.~\ref{figdosn4}.}\label{figdosn8}
\end{figure}
In Fig.~\ref{figdosn4} we consider two electrons on a ring of four sites. 
Comparing the running average (bold dot-dashed) over the discrete spectrum 
(thin dashed) and the classical limit (bold solid), we find good convergence 
already for a moderate spin length $S=10$. But even for the case $S=3/2$, 
which is realized in the manganites, the classical description appears to be 
acceptable. If we consider four electrons on a ring of eight sites, 
the spectrum is much more dense, making similarities to be recognized 
easily, see Fig.~\ref{figdosn8}. Of course, from the density of states 
we can learn nothing about correlations or other more involved features. 
Note, that in both figures we subtracted the peak at $E=0$ consuming a 
large fraction of spectral weight. 

\begin{figure}[htb]
  \epsfig{file=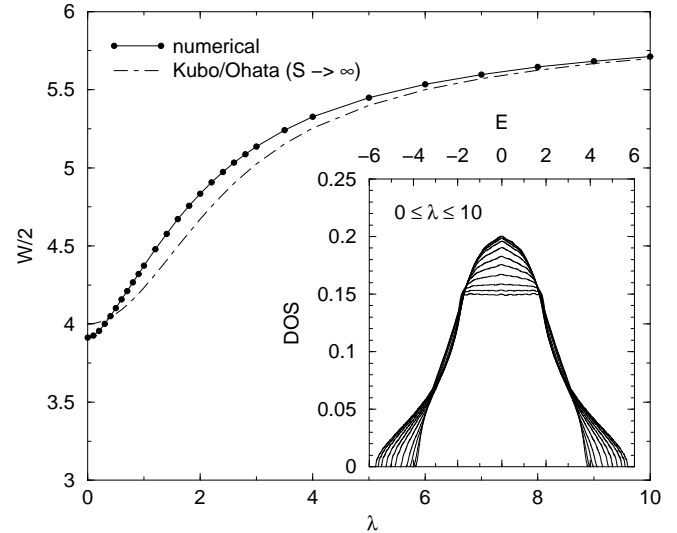, width=\linewidth}
  \caption{Bandwidth of tight-binding electrons on a simple cubic 
    lattice of size $64^3$, interacting with thermalized classical spins 
    in a homogeneous field $\lambda$, compared to the limit 
    $S\rightarrow\infty$ of the Kubo/Ohata formula;
    inset: Narrowing of the corresponding density of states with
    decreasing $\lambda$.}\label{figbwko}
\end{figure}

Another interesting check concerns the effective hopping matrix element 
$\tilde{t}^{\rm (b)}$, Eq.~(\ref{ttilde}). With classical spins and the 
Chebychev expansion methods mentioned above, one can calculate the 
grand-canonical DOS of the tight-binding model, Eq.~(\ref{hclass}), for
rather large clusters (here $64^3$ sites on a simple cubic lattice) 
without much effort. By considering a thermalized ensemble of classical 
spins in a homogeneous field we compare the resulting bandwidth with 
the limit $S\to\infty$ of $\tilde{t}^{\rm (b)}$, where the limit
of $\gamma_{\bar S}[\bar S \lambda]$ is obtained omitting $\bar S$ in
the argument and setting $\bar S\to\infty$ in the index,
\begin{equation}
  \gamma_{\bar S\to\infty}[\lambda] = \frac{1}{2} \left(
    1 + \coth(\lambda)\left[\coth(\lambda) - \frac{1}{\lambda}\right]
  \right)\,.
\end{equation}
The agreement is rather satisfactory, as can be seen in Fig.~\ref{figbwko}.
Naturally, the precise shape of the DOS (see inset) will not be reproduced
by the effective electronic model, it remains simple cubic tight-binding
for all fields and temperatures.

\end{document}